# GOOGLE SCHOLAR E ÍNDICE H EN BIOMEDICINA: LA POPULARIZACIÓN DE LA EVALUACIÓN BIBLIOMÉTRICA

**Álvaro Cabezas-Clavijo y Emilio Delgado-López-Cózar***

acabezasclavijo@gmail.com; edelgado@ugr.es

EC3: Evaluación de la Ciencia y de la Comunicación Científica, Departamento de Información y Comunicación, Facultad de Comunicación y Documentación, Universidad de Granada, Granada, España

*Autor para correspondencia

**RESUMEN:** El objetivo de este artículo es hacer una revisión de las características, prestaciones y limitaciones de los nuevos productos de evaluación científica derivados de Google Scholar, como son Google Scholar Metrics y Google Scholar Citations, y del índice h, el indicador bibliométrico adoptado como estándar por estos servicios. Asimismo se reseña la potencialidad de esta nueva base de datos como fuente para estudios en Biomedicina y se realiza una comparación del índice h obtenido por las revistas e investigadores más relevantes en el ámbito de la Medicina Intensiva, a partir de los datos extraídos de Web of Science, Scopus y Google Scholar. Los resultados muestran que, pese a que los valores medios de índice H en Google Scholar son casi un 30% más elevados que los obtenidos en Web of Science y en torno a un 15% más altos que los recogidos por Scopus, no hay variaciones sustantivas en las clasificaciones generadas a partir de una u otra fuente de datos. Aunque existen algunos problemas técnicos, se concluye que Google Scholar es una herramienta válida para los investigadores en Ciencias de la Salud, tanto a efectos de recuperación de información como de cara a la extracción de indicadores bibliométricos.

**PALABRAS CLAVE:** Evaluación de la investigación; Bibliometría; Índice h; Análisis de citas; Revistas científicas; Bases de datos bibliográficas; Google Scholar; Google Scholar Metrics; Google Scholar Citations; Biomedicina; Ciencias de la Salud.

**GOOGLE SCHOLAR AND THE H-INDEX IN BIOMEDICINE: THE POPULARIZATION OF BIBLIOMETRIC ASESSMENT**

**ABSTRACT:** The aim of this paper is to review the features, benefits and limitations of the new scientific evaluation products derived from Google Scholar; Google Scholar Metrics and Google Scholar Citations, as well as the h-index which is the standard bibliometric indicator adopted by these services. It also outlines the potential of this new database as a source for studies in Biomedicine and compares the h-index obtained by the most relevant journals and researchers in the field of Intensive Care Medicine, by means of data extracted from Web of Science, Scopus and Google Scholar. Results show that, although average h-index values in Google Scholar are almost 30% higher than those obtained in Web of Science and about 15% higher than those collected by Scopus, there are no substantive changes in the rankings generated from either data source. Despite some technical problems, it is concluded that Google Scholar is a valid tool for researchers in Health Sciences, both for purposes of information retrieval and computation of bibliometric indicators.

**KEYWORDS:** Research Evaluation; Bibliometrics; h-index; Citation Analysis; Periodicals; Databases, Bibliographic; Google Scholar; Google Scholar Metrics; Google Scholar Citations; Biomedicine; Health Sciences.

**Introducción**

Desde el punto de vista de la evaluación de la ciencia y de la comunicación científica el siglo XXI ha arrancado con el nacimiento de nuevas herramientas llamadas si no a sustituir, sí a complementar el uso de las tradicionales bases de datos de acceso a la información científica como Pubmed, Web of Science (WoS) o Scopus, y de los ya asentados indicadores bibliométricos con el factor de impacto a la cabeza. En este sentido, las nuevas herramientas de acceso y evaluación científica objeto de este ensayo (índice h y la familia Google Scholar) han supuesto la popularización del acceso a la información científica y de la evaluación de la investigación. Simples y sencillas de usar, calcular y entender -la divisa de estas nuevas herramientas-, así como de acceso libre y gratuito, han penetrado en el tejido científico a toda velocidad.

El objetivo de este artículo es hacer una revisión de las características de los nuevos sistemas de evaluación científica auspiciados por Google a partir de Google Scholar (Scholar Metrics, Scholar Citations) y del índice h, el indicador bibliométrico adoptado como estándar por estos servicios. Además, se realiza una comparación de los indicadores ofrecidos por estos productos con los que brindan los ya asentados Web of Science y Scopus, tomando como caso de estudio las revistas e investigadores más relevantes en el campo de la Medicina Intensiva.

**El índice h: un nuevo indicador bibliométrico para medir el rendimiento científico.**

El factor de impacto, concebido originalmente en los años 70 por Eugene Garfield, con la idea de seleccionar revistas para las colecciones bibliotecarias, bajo la asunción de que las revistas más citadas serían también las de mayor utilidad para los investigadores, y confeccionado a partir de los afamados Citation Indexes (Science Citation Index, Social Sciences Citation Index, hoy formando parte de la ISI Web of Science), sobrepasó con creces las expectativas de su creador y pronto se convirtió en el patrón oro de la evaluación, muy especialmente en el campo de la Biomedicina.

Aunque los abusos en la aplicación de este indicador han sido denunciados desde hace años (1,2), en los últimos tiempos han arreciado las críticas contra él, tanto por su forma de cálculo como por su aplicación indiscriminada en los procesos evaluativos (3,4,5), especialmente en los que conciernen al rendimiento individual de los investigadores, aspecto éste que es de todo punto incorrecto, como reconocía hace ya años el propio Eugene Garfield (6).

Es en ese contexto de búsqueda de indicadores válidos para medir el desempeño individual en el que nace el índice h. Ideado por Jorge Hirsch en agosto de 2005, la novedosa propuesta del físico argentino condensa en un solo indicador las dimensiones cuantitativa (producción) y cualitativa (citas) de la investigación, una vieja aspiración de la bibliometría. Un científico tiene un índice igual a h cuando h de sus artículos han recibido al menos h citas cada uno (7). Es decir, un investigador con índice h de 30 es aquel que ha conseguido publicar 30 artículos con al menos 30 citas cada uno de ellos.

El nuevo indicador despertó un inusitado interés entre los especialistas en Bibliometría, que pasaron a discutir sobre sus ventajas e inconvenientes y a ofrecer alternativas al mismo. Esto generó un auténtico aluvión de publicaciones, a tal punto que en 2011 se habían publicado ya más de 800 trabajos sobre el índice h[1]. Al margen de las numerosas variaciones propuestas para superar sus limitaciones (g, e, hc, hg, y, v, f, ch, hi, hm, AWCR, AWCRpA, AW…) (8), el índice h se ha extendido desde la evaluación de investigadores, su destino original, a la medición de instituciones (9), países (10) o revistas científicas (11,12).

El espaldarazo a la nueva métrica vino de la mano de las bases de datos centrales en Bibliometría, Web of Science y Scopus, que lo incorporaron inmediatamente a su batería de indicadores, facilitando la consulta a todos los niveles de agregación. Y su consagración definitiva quedó refrendada por Google, al adoptarlo en los dos productos de evaluación científica derivados de su buscador académico: Google Scholar Metrics, para revistas, y Google Scholar Citations, para investigadores.

El índice h ha quedado marcado en el subconsciente de muchos científicos como el indicador definitivo para medir el alcance de sus investigaciones. En líneas generales, esta métrica ha sido muy bien recibida por la comunidad científica por su facilidad de cálculo y por su capacidad para evaluar a los investigadores con un único número.

El índice h se caracteriza por su progresividad y robustez, así como por su tolerancia a los errores y desviaciones estadísticas. Se trata de un indicador progresivo, que sólo puede ir en aumento a lo largo de la carrera de un científico, creciendo exponencialmente la dificultad para alcanzar magnitudes más altas cuanto más alto sea el propio valor. Esto le confiere robustez, ya que los índices h de los investigadores crecerán de forma sostenida en el tiempo o se mantendrán estables, pero no experimentarán grandes variaciones, lo que permite valorar la trayectoria de un investigador más allá de puntuales trabajos altamente citados. Este mismo hecho le blinda frente a posibles errores en atribución de citas o búsquedas bibliográficas, habituales en las bases de datos comerciales.

Sin embargo, esta medida, al igual que todos los indicadores bibliométricos, no está exenta de algunas limitaciones:

- Para su cálculo adecuado debiéramos contar con un sistema de información o base de datos exhaustiva que contuviera todas las publicaciones y las citas asociadas a un investigador. Dado que con la sobreabundancia de información científica existente, esto es (por el momento) una entelequia, es fundamental tener presente que el indicador obtenido es directamente proporcional a la cobertura y calidad de la base de datos manejada. En este sentido, como se verá más adelante en este trabajo, los índices h variarán según la base de datos empleada (Google Scholar, WOS, Scopus).

---

[1] http://sci2s.ugr.es/hindex/biblio.php

- El índice h se verá afectado por la producción científica de la unidad estudiada (organización, revista, investigador). Las que cuentan con mayor producción científica pueden obtener mayores índices h, penalizando a los actores con estrategias de publicación selectivas (13).

- No es válido para comparar investigadores de áreas diversas, dado el distinto patrón de producción y citación de las disciplinas. Esta es una limitación inherente a todos los indicadores bibliométricos y sólo puede solucionarse adoptando el siguiente adagio: comparar lo comparable; jamás se deben comparar los índices h o los factores de impacto de disciplinas o especialidades diferentes ya que los tamaños de las comunidades científicas y el comportamiento en términos productivos y de citación de las mismas difieren en gran medida de una a otra. Como ejemplo baste decir que la revista con mayor impacto en Medicina Intensiva alcanza en 2011 un factor de impacto de 11,080 mientras que en Neurología la número uno presenta un índice de 30,445, sin que ello signifique en modo alguno que la segunda sea más relevante que la primera. Es conocido también que las revistas clínicas alcanzan cifras más modestas que las revistas de áreas básicas, y que las revisiones alcanzan un mayor número de citas que los artículos originales (2), por citar sólo algunas de las limitaciones del factor de impacto.

- No es adecuado para comparar investigadores noveles, de poca producción, con autores veteranos, de dilatada trayectoria científica. Para evitar este problema se sugiere comparar indicadores en ventanas de tiempo similares. Google Scholar ha adoptado esta perspectiva ofreciendo este indicador para los investigadores individuales, tanto para el conjunto de su carrera, como para el periodo más reciente, y en el caso de las revistas, para los últimos cinco años. Por su parte, tanto en Web of Science como en Scopus pueden acotarse con facilidad los períodos de tiempo de producción de las unidades estudiadas.

- No excluye las autocitas, limitación inseparable a los indicadores bibliométricos. Sin embargo, hay que señalar que la influencia de las autocitas sobre este indicador es menor que sobre otras métricas como el promedio de citas, ya que sólo afectaría a los documentos que estén en el límite de contar para el índice h (14). En cualquier caso, lo importante es que podamos calcular de forma rápida el indicador con y sin autocitas. Es lo que permiten hacer tanto WoS como Scopus, pero no Google Scholar.

- No es un indicador excesivamente discriminatorio, al tomar valores discretos y situarse en un rango habitualmente muy pequeño para la mayor parte de investigadores. Por ello es aconsejable acompañarlo de otras métricas que complementen la información que aporta el índice h.

- No tiene en cuenta la distinta responsabilidad y contribución de los autores en los trabajos, aspecto especialmente importante en biomedicina donde la investigación se hace en equipo lo que se refleja en un elevado número medio de autores por trabajo. En este sentido no es un indicador, como ocurre con el resto de los indicadores bibliométricos, que pondere la posición de los autores en la cadena de firmas, medio habitualmente empleado para identificar la atribución de responsabilidades en los trabajos

- No tiene en consideración los trabajos muy citados. Una vez superado el límite establecido por el índice h, las nuevas citas recibidas por un trabajo serán irrelevantes de cara a aumentar el h de un investigador. Por otra parte, como se ha comentado previamente, esta misma circunstancia le hace muy estable en el tiempo, evitando las desviaciones que dichos trabajos muy citados generan en otros indicadores de impacto.

Así, podemos concluir que el índice h es en esencia un indicador que, más que éxitos puntuales en la carrera científica, mide la regularidad de un investigador en su productividad e impacto científicos. Pese a su enorme facilidad de cálculo y uso, debe emplearse con suma precaución, ya que nunca es aconsejable medir el rendimiento de un investigador o institución con un solo indicador.

**La Familia Google: Google Scholar, Google Scholar Metrics y Google Scholar Citations.**

*Google Scholar*

Google Scholar, nacido en noviembre de 2004, es un buscador de documentos académicos (artículos en revistas, libros, *preprints*, tesis, comunicaciones a congresos, informes), que incluye información del número de veces que los trabajos han sido citados así como enlaces a los documentos citantes, ofreciendo acceso al texto completo de los trabajos, siempre que éstos estén disponibles libremente en la web, o se cuente con suscripción a ellos, si se accede desde una institución investigadora.

Gracias a su magnífico comportamiento en la recuperación de información, facilidad de uso, y diseño a imagen y semejanza del buscador general de Google, se ha convertido en parada obligada de un importante porcentaje de científicos a la hora de realizar búsquedas de información (15). Algunos estudios específicos en Biomedicina muestran como Google Scholar solo es superado en uso por Web of Science, y muestra prácticamente los mismos resultados que Pubmed (16). Este mismo estudio señala que Google Scholar se usa como fuente de datos complementaria a Web of Science o Pubmed, y que los usuarios valoran su facilidad de uso, velocidad en la respuesta y gratuidad, frente a la precisión y calidad en los resultados, que se perciben como factores determinantes en las otras dos fuentes de datos mencionadas. En el caso de Biomedicina varios estudios atestiguan la importancia de Google Scholar en su función de puerta de acceso a la información científica, como señalan tanto editoriales en las principales revistas biomédicas (17) como los estudios comparativos acerca de la exhaustividad y precisión en la recuperación de Google Scholar en comparación con Pubmed (18-21). Para el caso de España, y entre los médicos de atención primaria, González de Dios et al (22) documentó que el 70% de ellos accedía a la literatura científica a través de buscadores generales como Google o Yahoo y que tan sólo el 29% hacía uso de bases de datos especializadas.

Google Scholar permite acceder al texto completo de los trabajos académicos no sólo desde la fuente "oficial" sino también desde repositorios, webs personales y localizaciones alternativas, lo que proporciona enorme visibilidad a los materiales en acceso abierto. En este sentido, hay que señalar que la difusión en acceso abierto de materiales académicos es un gran aliado a la hora de facilitar la accesibilidad a las investigaciones e incrementar su difusión entre los investigadores del área, como señalan los propios datos en la especialidad de Medicina Intensiva (23).

Además, la ordenación de los resultados según relevancia (criterio que se deriva principalmente del número de citas recibidas por los trabajos), asegura que los resultados no sólo son pertinentes a nuestra búsqueda, sino también relevantes científicamente, al menos en lo que respecta a los primeros registros devueltos por el buscador. Habitualmente, dichos resultados son artículos altamente citados, que provienen de publicaciones de prestigio y bien conocidas por los investigadores, por lo que a la rapidez y facilidad en la recuperación de información se le suma el hecho de que el investigador se siente "cómodo" con los resultados que obtiene. Como demuestra un reciente estudio realizado sobre los veinte primeros resultados de diversas búsquedas clínicas, los artículos recuperados por Google Scholar reciben un número de citas mayor y se publican en revistas de mayor factor de impacto que los documentos recuperados por Pubmed (21).

Sin embargo, Google Scholar presenta una serie de limitaciones que han de ser tenidas en cuenta a la hora de su utilización como fuente de información (24,25). La principal de estas barreras es la política de inclusión de fuentes del buscador, que es opaca. Pese a que es conocido que Google Scholar rastrea la información contenida en sitios webs de universidades y centros de investigación, repositorios, o sitios webs de editores científicos, no sabemos exactamente qué fuentes son esas, ni por supuesto el tamaño de la base documental. Por otra parte, junto a trabajos de investigación, es posible también encontrar materiales docentes, e incluso de carácter administrativo, poco relevantes de cara a la labor investigadora. Asimismo, los datos de Google Scholar no tienen ningún tipo de normalización, consecuencia de la amplia cobertura, la variedad de fuentes de información y el procesamiento automático de la información (25).

Los estudios comparativos entre Google Scholar y Pubmed, el verdadero sancta sanctorum de las bases de datos bibliográficas en Biomedicina proporcionan resultados muy similares en lo que respecta a exhaustividad en la recuperación de información, si bien resultando más favorables al producto de la National Library of Medicine estadounidense cuando se trata de definir búsquedas de cierta complejidad (20). Es por ello que puede ser útil emplear una estrategia híbrida de cara a la recuperación documental, usando Pubmed cuando se trate de realizar la búsqueda bibliográfica previa a cualquier estudio (18), y Google Scholar para identificar y acceder de forma rápida a documentos concretos, o en búsquedas muy definidas como artículos sobre un tema en una revista dada, o de un autor determinado.

Su característica de recoger las citas a los trabajos indizados le ha situado como un serio competidor de las bases de datos tradicionales usadas para la evaluación de la ciencia, generándose una amplia literatura que compara resultados a partir de los datos de Web of Science, Scopus y Google Scholar (26-28). Autores como Harzing y Der Wal (29) han llegado a afirmar que Google Scholar supone una *democratización* del análisis de citas.

### *Google Scholar Citations: perfiles bibliométricos de investigadores*

Sabedor de la rápida expansión de Google Scholar entre los científicos, y conocedor de las necesidades que este sector tiene desde el punto de vista de la evaluación científica, Google se aprestó en julio de 2011 a lanzar un nuevo producto. Google Scholar Citations (Mis citas, en español), es un servicio que permite a un investigador establecer automáticamente una página con su perfil científico, donde figuran los documentos publicados y recogidos en Google Scholar, así como el número de citas que cada uno de ellos ha recibido, generando una serie de indicadores bibliométricos, encabezados por el omnipresente índice h (30). Estas estadísticas se renuevan automáticamente con la nueva información localizada por Google Scholar (nuevas publicaciones de las que somos autores o nuevas citas a nuestros trabajos), por lo que el perfil científico está permanentemente actualizado sin que el autor tenga que hacer nada. Scholar Citations presenta tres indicadores bibliométricos. Además del mencionado índice h, señala el número total de citas de los trabajos de un investigador y el índice i10, esto es, el número de trabajos del investigador con más de diez citas (figura 1). Los mencionados indicadores se calculan tanto para el conjunto de la carrera académica como para el período más reciente (seis últimos años).

**Figura 1** Perfil de un investigador en Google Scholar Citations.

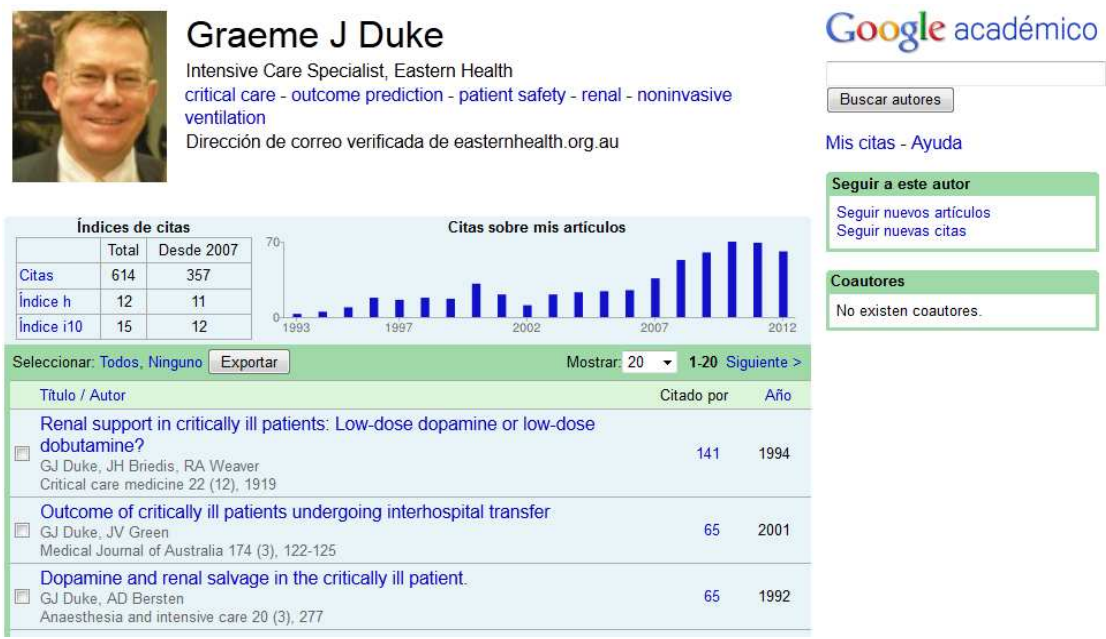

El proceso de alta de un investigador es realmente sencillo; una vez proporcionada una dirección de correo institucional, y los nombres de firma (y sus variantes) que solemos usar, Google Scholar nos señalará los trabajos de los que somos autores con el objeto de que los confirmemos o rechacemos. Esos trabajos componen el corpus documental a partir del cual se obtendrán nuestros datos bibliométricos. Además el investigador puede optar entre hacer público o no su perfil, así como editar los registros, con lo que se consigue corregir o normalizar la información del buscador, unir registros duplicados e incluso añadir de forma manual otros trabajos que no figuren en el buscador académico. En este último caso, sin embargo, no contaremos con el número de citas recibidas por dichos documentos. Asimismo, el sistema permite exportar los registros en los formatos bibliográficos más habituales.

Otra ventaja de establecer un perfil en Scholar Citations es que recibiremos actualizaciones de documentos relevantes a nuestros intereses académicos. Por medio de algoritmos que tienen en cuenta las palabras que usamos en nuestros trabajos y los coautores de los mismos, así como los flujos de citas entre artículos y revistas, Google Scholar nos ofrecerá documentos adecuados a nuestras necesidades científicas.

Asimismo podemos vincularnos a coautores que estén registrados en Scholar Citations con perfil público, y añadir las palabras clave o descriptores que mejor definan nuestro trabajo. Esta prestación es digna de mención, ya que es el propio investigador quien indica las disciplinas o subdisciplinas en las que es experto. Con ello se posibilita la creación de un listado de investigadores según el número total de citas recibidas (figura 2). En este sentido hay que señalar la lamentable falta de normalización en cuanto a la asignación de descriptores, y que posibilita que un mismo concepto se represente de diversas formas. Por ejemplo, en el campo de Medicina Intensiva podemos encontrar los descriptores *Critical Care*, *Critical Care Medicine*, *Intensive Care* o *Medicina Intensiva*, entre otros. Igualmente se ha detectado que algunos investigadores usan términos poco comunes a fin de aparecer como los más destacados en una subespecialidad concreta. Otro fenómeno es el de investigadores con actividad en más de un campo del conocimiento. Por ejemplo, es frecuente en Bibliometría la aparición de científicos de campos como Física, Química o Informática, que les posibilita aparecer en los primeros lugares en disciplinas que no constituyen el grueso de su actividad, originando listados poco realistas, o directamente falsos. En las disciplinas en las que los autores de este trabajo son especialistas (*Bibliometrics*, *Scientometrics*) es notoria la presencia junto a destacados investigadores de la disciplina, de científicos absolutamente desconocidos para los especialistas en el campo. En este sentido cabría esperar una mayor seriedad por parte de Google al unir los descriptores que puedan considerase cuasi sinónimos, incluso se podría sugerir el uso de vocabularios normalizados como los encabezamientos MESH en Biomedicina. Por su parte, en el caso de los investigadores, sería deseable un mayor rigor a la hora de vincularse a las disciplinas.

**Figura 2** Investigadores con mayor número de citas en Google Scholar Citations vinculados al descriptor Medicina Intensiva (Critical Care).

Desde un punto de vista bibliométrico, se detectan algunas limitaciones en este servicio. Por ejemplo, la ausencia del número total de publicaciones de un investigador, el indicador más evidente para tomar el pulso de la actividad científica de un investigador es la falta más notoria. De igual manera se echan en falta indicadores vinculados a la capacidad para publicar en revistas o medios de difusión de prestigio, si bien puede que la introducción de Scholar Metrics pueda suplir esta carencia en el corto plazo, en el caso de que Google se decida a integrar ambos productos.

Otro aspecto a contemplar es la posible picaresca de los autores. Al igual que se ha detectado en el caso de las disciplinas y subdisciplinas a las que uno se adscribe, es factible que un investigador se asigne artículos que no le corresponden, como bien señala Anne W. Harzing (31), si bien estas prácticas son fácilmente detectables. Otra posibilidad es la manipulación de la base de datos introduciendo autores falsos que inflen nuestros registros de citas. Los experimentos de Labbé (32) y Delgado López-Cózar et al (33) demuestran que esto es factible y sencillo, lo que le diferencia de entornos cerrados y controlados como los de las grandes bases de datos Web of Science o Scopus, donde dicha manipulación, si bien posible, es mucho más compleja.

Otra duda es si los directorios generados serán representativos de la investigación en un campo. Los investigadores con escaso impacto académico disponen de pocos incentivos para crear su perfil público, ya que podrían quedar en evidencia respecto a otros compañeros con currículos de mayor nivel. Así pues, uno de los riesgos es crear un directorio de científicos de alto impacto, poco representativo de la investigación que se da en un campo (30).

*Google Scholar Metrics: rankings de revistas científicas*

En abril de 2012, Google lanzó Google Scholar Metrics (GSM), denominado en español Estadísticas. Se trata de un producto bibliométrico, gratuito y de libre acceso, que ofrece el impacto de las revistas científicas y de otras fuentes documentales, y que usa el índice h como criterio de ordenación.

La consulta de este producto se puede realizar de dos maneras: la primera de ellas es acceder a los rankings de las cien revistas con mayor índice h según su idioma de publicación (figura 3). En la actualidad son diez los idiomas disponibles: inglés, chino, portugués, alemán, español, francés, coreano, japonés, holandés e italiano, si bien a este respecto hay que señalar que GSM no parece diferenciar bien las revistas con contenidos bilingües. Desde noviembre de 2012, fecha en que se lanzó la última actualización del producto, se pueden consultar también las primeras veinte publicaciones según su índice h para ocho áreas del conocimiento, y 313 disciplinas, si bien sólo para las revistas en inglés (34).

**Figura 3** Listado de publicaciones con mayor índice h en español (2007-2011) según Google Scholar Metrics.

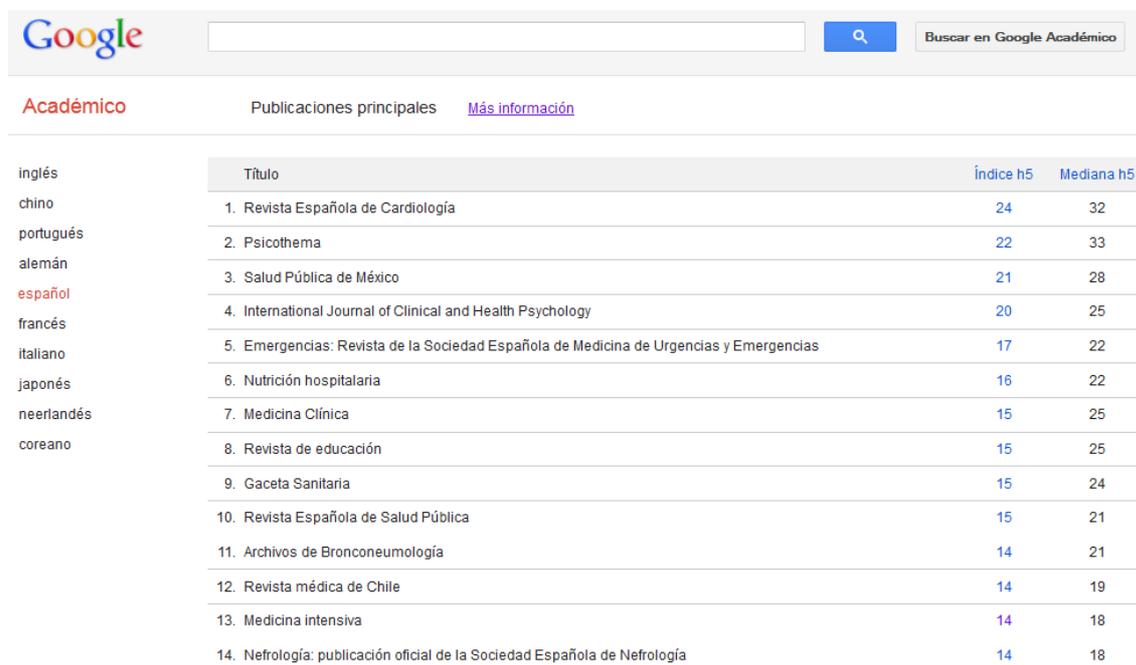

La segunda opción es insertar directamente en el cuadro de búsqueda palabras incluidas en los títulos de las revistas. En este caso, la búsqueda se realiza sobre todas las fuentes indizadas por GSM (no tienen por qué estar entre el top de cada idioma o disciplina), devolviendo un máximo de veinte resultados ordenados según índice h. Las fuentes indizadas en GSM son aquellas que han publicado al menos 100 artículos en el período 2007-2011 y recibido alguna cita (esto es, se excluyen las que tienen un índice h=0). Este indicador cubre los trabajos publicados entre 2007 y 2011, así como las citas recibidas hasta el 15 de noviembre de 2012. De este modo, es un sistema de información estático, con actualizaciones periódicas, si bien Google no desvela con qué carácter se seguirán realizando éstas.

Para cada una de las revistas se puede acceder a la información sobre los artículos que contribuyen al índice h. Así para una revista con un índice h de 14, como es el caso de Medicina Intensiva, se muestran los 14 artículos que han recibido 14 o más citas (figura 4). Del mismo modo, pinchando sobre ellos, se puede acceder a los documentos citantes, de modo que se puede analizar con facilidad las fuentes que contribuyen en mayor medida al índice h de una revista.

**Figura 4** Artículos más citados de Medicina Intensiva (2007-2011) según Google Scholar Metrics.

El hecho de que se usen marcos temporales bastante amplios (cinco años) asegura que este indicador se mantendrá bastante estable a lo largo del tiempo. Dado que el índice h es un indicador con poco poder discriminatorio (toma valores discretos y es muy difícil aumentarlo a partir de cierto umbral de citación), Google ha incluido otro indicador, la mediana de las citas recibidas por los artículos que contribuyen al índice h, como segundo criterio para ordenar las revistas.

Sin embargo, Google Scholar Metrics muestra una serie de limitaciones muy importantes, que lo muestran poco fiable de cara a la evaluación científica (34,35):

- Cobertura: mezcla de indiscriminada en los rankings de fuentes documentales tan distintas como revistas, series y colecciones incluidas en repositorios, bases de datos o actas de congreso.

- Errores en los recuentos de citas por falta de normalización de títulos de revistas, especialmente en aquellas revistas con versiones en varios idiomas.

- Presentación principal de resultados por idiomas y no por disciplinas científicas, que es lo habitual en productos de este tipo. Se comparan de este modo revistas de disciplinas científicas distintas, con distintos patrones de producción y citación.

- Imposibilidad de buscar por áreas o disciplinas en las revistas en idiomas distintos del inglés. Limitación de resultados en la búsqueda por palabras clave a los 20 títulos con mayor índice h.

- Ventana de citación insuficiente para evaluar publicaciones de alcance nacional en disciplinas de Ciencias Sociales y Humanas

Señaladas estas limitaciones, hay que valorar positivamente este paso adelante de Google, queva a facilitar la consulta del impacto de las revistas por parte de investigadores sin acceso a las tradicionales bases de datos de citas, de importante coste económico. Este hecho, además de acercar los indicadores bibliométricos a todos los investigadores, puede estimular la competencia entre los diferentes productos.

**La Medicina Intensiva a través del índice h: comparando Google Scholar, Web of Science y Scopus**

A fin de comprobar el comportamiento de Google Scholar y Google Scholar Metrics, y visualizar claramente las diferencias en los resultados en comparación con las bases de datos tradicionales, se han contrastado los índices h obtenidos en una muestra de revistas y de autores significativos dentro de la Medicina Intensiva. Como medida de comparación de las posiciones de revistas y autores, se ha usado el coeficiente de correlación de Spearman (rho), estadístico usado habitualmente en los estudios bibliométricos para medir el grado de asociación entre dos variables a partir de su posición en diferentes rankings (36).

Hemos realizado el ejercicio comparativo con las 26 revistas recogidas en los Journal Citations Reports (JCR) en 2011 dentro de la categoría Medicina Intensiva (Critical Care Medicine) en tres fuentes de datos, Web of Science (WoS), Scopus y Google Scholar Metrics (GSM). Tomando los artículos publicados entre 2007 y 2011, las revistas American Journal of Respiratory and Critical Care Medicine, Chest y Critical Care Medicine muestran los índices h más elevados, independientemente de la fuente de datos usada. El valor medio para el conjunto de las 26 revistas en este indicador es de 28 para el caso de WoS, 32 para las revistas en Scopus, y 36 en GSM (tabla 1).

Esto significa que los datos de GSM son un 28% superior a los obtenidos en WoS, y un 13% superior a los recogidos en Scopus. Hay que señalar que los datos de GSM contabilizan las citas recibidas hasta abril de 2012, mientras que el cálculo efectuado sobre las otras dos bases de datos se realizó seis meses más tarde, por lo que la diferencia porcentual entre GSM y las otras dos bases de datos puede estar algo infravalorada.

**Tabla 1** Índice h de las revistas de Medicina Intensiva (n=26) en Web of Science, Scopus y Google Scholar Metrics (2007-2011)

| REVISTA | ÍNDICE H | | | POSICIÓN | | |
|---|---|---|---|---|---|---|
| | WOS | SCOPUS | GSM | WOS | SCOPUS | GSM |
| American Journal of Respiratory and Critical Care Medicine | 81 | 93 | 96 | 1 | 1 | 1 |
| Chest | 68 | 79 | 89 | 2 | 2 | 2 |
| Critical Care Medicine | 64 | 74 | 83 | 3 | 3 | 3 |
| Intensive Care Medicine | 47 | 55 | 63 | 4 | 4 | 4 |
| Journal of Trauma-Injury Infection and Critical Care | 47 | 40 | 53 | 4 | 6 | 5 |
| Critical Care | 40 | 45 | 53 | 6 | 5 | 5 |
| Journal of Neurotrauma | 37 | 40 | 40 | 7 | 6 | 8 |
| Resuscitation | 36 | 40 | 45 | 8 | 6 | 7 |
| Shock | 31 | 34 | 38 | 9 | 9 | 9 |
| Injury | 29 | 34 | 38 | 10 | 9 | 9 |
| Current Opinion in Critical Care | 24 | 27 | 32 | 11 | 11 | 11 |
| Pediatric Critical Care Medicine | 23 | 27 | 30 | 12 | 11 | 12 |
| Burns | 23 | 25 | 29 | 12 | 13 | 13 |
| Neurocritical Care | 21 | 24 | 25 | 14 | 14 | 15 |
| Journal of Critical Care | 20 | 23 | 26 | 15 | 15 | 14 |
| Seminars in Respiratory and Critical Care Medicine | 19 | 23 | sd | 16 | 15 | sd |
| Respiratory Care | 19 | 22 | 23 | 16 | 17 | 17 |
| American Journal of Critical Care | 17 | 21 | 25 | 18 | 18 | 15 |
| Critical Care Clinics | 17 | 19 | 22 | 18 | 19 | 18 |
| Minerva Anestesiologica | 17 | 19 | 19 | 19 | 19 | 19 |
| Anaesthesia and Intensive Care | 14 | 17 | 18 | 21 | 21 | 20 |
| Critical Care Nurse | 9 | 13 | 18 | 22 | 22 | 20 |
| Critical Care and Resuscitation | 8 | 12 | 12 | 23 | 23 | 23 |
| Anasthesiologie und Intensivmedizin | 8 | 8 | 4 | 23 | 25 | 25 |
| Medicina Intensiva | 7 | 10 | 14 | 24 | 24 | 22 |
| Anästhesiologie, Intensivmedizin, Notfallmedizin, Schmerztherapie | 6 | 7 | 7 | 25 | 26 | 24 |
| **PROMEDIO** | 28 | 32 | 36 | | | |

Datos de GSM tomados de http://scholar.google.com/citations?view_op=top_venues (número de citas actualizado a 1 de abril de 2012). Datos de Web of Science y Scopus recopilados el 26 de septiembre de 2012.

Por su parte, el promedio de índice h de la muestra de revistas de Medicina Intensiva analizada es en Scopus un 13,5% superior al obtenido en WOS (tabla 2). Sin embargo, la comparación del orden de las revistas según su índice h realizada a través de la correlación de Spearman (Rho) muestra que las diferencias son inapreciables entre bases de datos. El valor de la correlación, 0,99 en los tres casos, a un nivel de significación del 1% señala que pese a las diferencias cuantitativas en cuanto al índice h entre bases de datos, la ordenación de las revistas es prácticamente la misma independientemente de la fuente de datos que utilicemos.

**Tabla 2** Comparación del índice h de revistas de Medicina Intensiva según bases de datos

| FUENTE | N† | Rho* | DIFERENCIA H-INDEX |
|---|---|---|---|
| WOS-SCOPUS | 26 | 0,994 | +13,5% (SCOPUS) |
| WOS-GSM | 25 | 0,989 | +28,2% (GSM) |
| SCOPUS-GSM | 25 | 0,994 | +12,9% (GSM) |

N=revistas; Rho= coeficiente de correlación de Spearman; Diferencia h-Index: diferencias porcentuales de los promedios de h-index entre bases de datos. Entre paréntesis, base de datos con mayor promedio de índice h. *La correlación es significativa al 1%. †Al existir una revista en GSM para la que no se cuentan con datos, el análisis estadístico se realiza sobre 25 revistas en el caso de esta base de datos, en lugar de sobre las 26 revistas sobre las que se realiza el análisis estadístico en la comparación WOS-Scopus.

Al igual que para las revistas, se ha calibrado el índice h de científicos destacados en el campo de la Medicina Intensiva. Se han tomado los 20 investigadores más productivos en 2007-2011 dentro de la categoría Medicina Intensiva (Critical Care Medicine) en Web of Science. Para ello, se diseñó la siguiente ecuación de búsqueda a través de la búsqueda avanzada (WC=Critical Care Medicine AND PY=2007-2011). De este modo, y haciendo uso de la opción Refine Results de la propia base de datos, se tomaron los investigadores más productivos, calculándose su índice h tanto en Web of Science, como en Scopus y en Google Scholar. De esta muestra se excluyeron algunos investigadores con nombres comunes, que podían entorpecer el proceso de búsqueda. Los datos de los índices h de estos investigadores pueden consultarse en la tabla 3. El valor medio de índice h asciende a 23 en el caso de Web of Science, 25 para Scopus, y 29 si se toma Google Scholar como fuente para el análisis.

**Tabla 3** Índice h de investigadores destacados en Medicina Intensiva (n=20) en Web of Science, Scopus y Google Scholar (2007-2011)

| | ÍNDICE H | | | POSICIÓN | | |
|---|---|---|---|---|---|---|
| INVESTIGADOR | WOS | SCOPUS | GS | WOS | SCOPUS | GS |
| HOLCOMB JB | 37 | 36 | 43 | 1 | 2 | 2 |
| BELLOMO R | 36 | 40 | 48 | 2 | 1 | 1 |
| MATTHAY MA | 27 | 29 | 35 | 3 | 3 | 3 |
| RELLO J | 25 | 27 | 33 | 4 | 5 | 4 |
| ANGUS DC | 25 | 26 | 31 | 4 | 8 | 6 |
| REINHART K | 25 | 26 | 31 | 4 | 8 | 6 |
| MOORE EE | 24 | 27 | 31 | 7 | 5 | 6 |
| KOLLEF MH | 24 | 27 | 31 | 7 | 5 | 6 |
| AZOULAY E | 23 | 26 | 28 | 9 | 8 | 11 |
| WADE CE | 23 | 25 | 29 | 9 | 11 | 10 |
| KELLUM JA | 22 | 28 | 33 | 11 | 4 | 4 |
| ELY EW | 21 | 24 | 25 | 12 | 12 | 14 |
| SCALEA TM | 21 | 22 | 25 | 12 | 14 | 14 |
| HERNDON DN | 20 | 22 | 28 | 14 | 14 | 11 |
| GIANNOUDIS PV | 19 | 24 | 27 | 15 | 12 | 13 |
| ANNANE D | 19 | 21 | 25 | 15 | 16 | 14 |
| GAJIC O | 18 | 20 | 22 | 17 | 17 | 18 |
| DEMETRIADES D | 18 | 19 | 23 | 17 | 18 | 17 |
| BROCHARD L | 15 | 19 | 21 | 19 | 18 | 19 |
| SPRUNG CL | 15 | 19 | 20 | 19 | 18 | 20 |
| **PROMEDIO** | 23 | 25 | 29 | | | |

Datos recopilados el 28 de septiembre de 2012.

Las diferencias en cuanto al valor promedio de índice h entre las bases de datos son muy similares a las halladas para las revistas, obteniéndose en Google Scholar valores casi un 30% por encima de lo encontrado en Web of Science, y un 16% por encima de Scopus. Por su parte, la diferencia entre Web of Science y Scopus, a favor de ésta última es apenas del 10,9% (tabla 4).

**Tabla 4** Comparación del índice h de investigadores en Medicina Intensiva según bases de datos

| FUENTE | N | Rho* | DIFERENCIA H-INDEX |
|---|---|---|---|
| WOS-SCOPUS | 20 | 0,918 | +10,9% (SCOPUS) |
| WOS-GS | 20 | 0,934 | +28,9% (GS) |
| SCOPUS-GS | 20 | 0,968 | +16,2% (GS) |

N=investigadores; Rho= coeficiente de correlación de Spearman; Diferencia h-Index: diferencias porcentuales de los promedios de h-index entre bases de datos. Entre paréntesis, base de datos con mayor promedio de índice h. *La correlación es significativa al 1%.

Se observan escasas diferencias en los rankings en función de una u otra variable, si bien el investigador más relevante en Web of Science (Holcomb JB) no lo es en Scopus ni en Google Scholar (Bellomo R). El valor de la correlación de Spearman (Rho) muestra que ésta es muy alta entre las tres fuentes de datos, especialmente entre Scopus y Google Scholar (0,968), a un nivel de significación del 1%.

**Discusión**

En la actualidad existen multitud de formas de evaluar la investigación bajo parámetros bibliométricos, y diversas fuentes de información para ello, con lo que, de facto, se ha superado el monopolio del factor de impacto y de ISI Web of Science para dichas tareas evaluadoras. Indicadores como el índice h o bases de datos y productos como los aportados por Google son sólo algunas de las expresiones más reconocibles de dicha labor evaluadora.

El índice h es ante todo un indicador a emplear para valorar la trayectoria científica de un investigador, midiendo su regularidad en cuanto a su productividad e impacto científicos. Asimismo, puede emplearse para medir el rendimiento de las revistas e instituciones, si bien en este caso es recomendable limitar su cálculo a períodos limitados de tiempo. Pese a su enorme facilidad de cálculo y uso, y a su robustez y tolerancia a las desviaciones estadísticas debe emplearse con suma precaución y, en cualquier caso, deben imponerse medidas y controles que neutralicen o minimicen sus sesgos. Uno de estos controles, en el caso de la Biomedicina, sería realizar las comparaciones siempre dentro de una misma disciplina, y en el caso de los investigadores, ponderar carreras científicas de parecida duración o bien aplicar algún criterio de normalización en función de los años de trayectoria investigadora o de la posición académica que se ocupe. Asimismo hay que enfatizar que nunca es aconsejable medir el rendimiento de un investigador o institución con una única métrica, por lo que en todo momento se recomienda valerse de una batería de indicadores en los procesos evaluativos bajo parámetros bibliométricos, y a ser posible en conjunción con el asesoramiento experto de los especialistas del campo científico correspondiente.

Ahora bien, debe quedar claro que el empleo de índice h no excluye que el factor de impacto deba seguir usándose en este campo como indicador de calidad científica. Es un buen termómetro para medir las dificultades y barreras para acceder a una determinada revista científica, así como para evaluar la repercusión de ésta. En el caso de la Medicina es requisito inexcusable para los investigadores publicar sus trabajos en revistas que cuenten con factor de impacto, ya que éste sigue siendo el patrón oro para evaluar el desempeño de las revistas científicas (37). Las revistas con mayores factores de impacto reciben un gran número de manuscritos, viéndose obligadas a rechazar una gran parte, por lo que los autores deben competir entre sí para acabar publicando en ellas. Es por ello que el factor de impacto se puede considerar como un indicador de competitividad (38). Baste el ejemplo de esta misma revista que aumentó en un 200% el número de originales recibidos tras obtener su primer factor de impacto (39). Este fenómeno señala que es la propia comunidad científica la que se auto-regula, ya que las revistas de mayor difusión incorporan a su vez controles más exhaustivos y exigentes, permitiendo sólo la publicación a los miembros más destacados de la comunidad científica. Lógicamente, esto no evita que en estas revistas también se publiquen junto a investigaciones relevantes, trabajos de escasa valía o utilidad para los investigadores, pero estos trabajos serán un porcentaje significativamente menor al que se difunden a través de revistas sin factor de impacto. Por esto conviene también recordar que el factor impacto de las revistas no es representativo del impacto de los artículos publicados en ellas. La práctica de considerar que el trabajo publicado en una revista, el autor que lo firma y la institución que lo cobija heredan mecánicamente el valor del factor de impacto de la revista donde se ha publicado es un mal muy extendido, a pesar de lo erróneo de este planteamiento, muy especialmente en el caso de los investigadores españoles (40).

En definitiva, es importante contemplar el hecho de que todos los indicadores bibliométricos poseen limitaciones, y que no se debe usar un criterio único en la evaluación de la investigación, sino que es necesario aplicar un conjunto de indicadores a la hora de realizar valoraciones rigurosas y justas.

En lo que concierne a Google Scholar como fuente de información científica cabe concluir que es una seria alternativa a las tradicionales bases de datos. El buscador de Google ofrece resultados similares a los de las otras fuentes de datos, y puede ser preferible su uso para una primera aproximación documental a una temática concreta o para la localización de determinadas referencias bibliográficas. Sin embargo, en Biomedicina sigue siendo preferible el entorno controlado que aportan fuentes de datos especializadas como Pubmed, con revistas y contenidos validados por la comunidad científica.

En cuanto a Google Scholar Citations y Google Scholar Metrics, los productos que se derivan de Google Scholar, si bien arrastran los problemas de falta de normalización del buscador, proporcionan de forma rápida y gratuita información bibliométrica muy valiosa tanto del desempeño de investigadores individuales como de revistas científicas, lo que sin duda contribuye a familiarizar a los investigadores con los indicadores bibliométricos y con su uso en la evaluación de la actividad científica.

Ahora bien lo que resulta sorprendente cuando comparamos los resultados bibliométricos ofrecidos por Google Scholar y las tradicionales bases de datos bibliométricas Web of Science y Scopus, tanto si evaluamos investigadores como revistas es que son prácticamente idénticos, por lo menos en lo que atañe a la Medicina Intensiva. No hay disparidades en los rankings obtenidos: los tres sistemas coinciden en identificar a los autores y revistas más relevantes e influyentes. Las únicas diferencias son de escala y tamaño: Google Scholar recupera más citas que Scopus y éste, a su vez, más que Web of Science.

Bien es verdad, que, más allá de los aspectos técnicos, hay dos diferencias importantes. La primera es de orden conceptual. El *ethos* científico exige controles previos que certifiquen y validen la generación y difusión del conocimiento, esto es, exige medios controlados. Pues bien, Google Scholar trabaja con un entorno incontrolado ya que procesa toda la información científica que es capaz de capturar en la Internet académica sin más, independientemente de cómo se produzca y comunique. En este sentido, las nuevas fuentes de información dinamitan y se enfrentan a las bases de datos tradicionales que se alimentan exclusivamente de fuentes científicas filtradas y depuradas (revistas y congresos) con los más rigurosos controles científicos. Sin embargo, a efectos de evaluación científica da igual el control que el "descontrol". La segunda diferencia es de orden crematístico, pues no es lo mismo el precio a pagar por unas fuentes u otras. Mientras Google Scholar es gratuita y de libre uso, las bases de datos multidisciplinares suponen a las arcas de las instituciones de investigación millonarias facturas, además de imponer licencias muy restrictivas.

En conclusión, si el factor de impacto significó la entrada de la evaluación del rendimiento científico en la morada del mundo académico, y Google la popularización del acceso a la información científica, el índice h y las herramientas bibliométricas gestadas a partir de Google Scholar han significado la extensión de dicha evaluación científica, haciendo más accesible la recuperación de información académica y el cómputo de indicadores bibliométricos. En definitiva, la gran novedad de estos últimos años es que dichos productos han dejado de ser coto exclusivo de especialistas y están ahora, gracias a la facilidad de cálculo de estas métricas y de la gratuidad y accesibilidad de las nuevas herramientas, al alcance de cualquier investigador, en una suerte de popularización de la evaluación científica.

**Financiación**

**Bibliografía**